\documentclass[12pt,a4paper]{article}

\usepackage{geometry}
\usepackage{graphicx}
\usepackage{pdflscape}
\usepackage{caption}
\usepackage{subcaption}
\usepackage{float}
\usepackage{url}
\usepackage{listings}
\usepackage{color}
\usepackage{fullpage}
\usepackage{booktabs}
\usepackage{array}
\usepackage{siunitx}

\newcolumntype{R}[1]{>{\raggedleft\let\newline\\\arraybackslash\hspace{0pt}}m{#1}}

\title{Pheet meets C++11}
\author{Manuel P\"oter
\thanks{This report was the outcome of a project from autumn 2013 to
summer 2014 at the Parallel Computing group at the
Vienna University of Technology. The project was supervised by Martin Wimmer 
and Jesper Larsson Tr\"aff, and also resulted in~\cite{Wimmer:2013:PTF:2510648.2510846}.}
}

\begin{document}

\maketitle

\definecolor{mygreen}{rgb}{0,0.5,0}
\definecolor{mygray}{rgb}{0.5,0.5,0.5}
\definecolor{mymauve}{rgb}{0.58,0,0.82}
\lstset{
  backgroundcolor=\color{white},
  basicstyle=\footnotesize,
  breakatwhitespace=false,
  breaklines=true,
  captionpos=t,
  commentstyle=\color{mygreen},
  deletekeywords={...},
  escapeinside={\%*}{*)},
  extendedchars=true,
  frame=bt,
  keepspaces=true,
  keywordstyle=\color{blue},
  morekeywords={*,...},
  numbers=left,
  numbersep=5pt,
  numberstyle=\color{mygray},
  rulecolor=\color{black},
  showspaces=false,
  showstringspaces=false,
  showtabs=false,
  stepnumber=1,
  stringstyle=\color{mymauve},
  tabsize=2,
  language=C++
}

\begin{abstract}
Pheet is a C++ task-scheduling framework that allows for easy customization of internal data-structures. The implementation was started before the C++11 standard was committed and therefore did not use the new standardized memory model but compiler/platform specific intrinsics for atomic memory operations. This not only makes the implementation harder to port to other compilers or architectures but also suffers from the fact that prior C++ versions did not specify any memory model.

In this report I discuss the porting of one of the internal Pheet data structures to the new memory model and provide reasoning about the correctness based on the semantics of the memory consistency model. Using two benchmarks from the Pheet benchmark suite I compare the performance of the original against the new implementation which shows a significant speedup under certain conditions on one of the two test machines.
\end{abstract}

\section*{Introduction}
Pheet\footnote{\url{http://www.pheet.org}}, developed by Wimmer \cite{Wimmer:2014:diss}, is a highly customizable task scheduling framework developed in C++ that allows comparison between different implementations of data structures used in the scheduler, as well as comparisons between entirely different schedulers. The customization options are based on the powerful template meta-programming possibilities provided by the C++ language. Even though the framework is fully customizable and allows replacing almost all of the internal data structures, the template meta-programming approach produces high-performance code since C++ templates are instantiated at compile time and therefore all the usual compiler optimizations like inline expansion, dead-code elimination, etc. can by applied. It is easy to use and provides good performance. 

Although the framework makes extensive use of many new language features that were introduced in C++11 (mainly template aliases and variadic templates) the development started before the C++11 standard was officially adopted by the standards committee. For this reason the original implementation did not use the new memory model that was introduced with C++11, but instead relied on macros that resolved to compiler/platform specific intrinsics.

The C++ Standard prior to C++11 specified program execution in terms of observable behavior, which in turn described sequential execution on an implicitly single-threaded abstract machine. Therefore multi-threaded C++ programs relied on set of libraries for threading support like POSIX threads, Win32, or Boost. Unfortunately a pure library approach, in which the compiler is
designed independently of threading issues, includes all sorts of problems \cite{Boehm:2005:TCI:1064978.1065042}. Without a clearly defined memory model as a common ground between the compiler, the hardware, the threading library, and the programmer, multi-threaded C++ code is fundamentally at odds with compiler and processor-level optimizations \cite{Meyers2004}.

In August 2011 the ISO C++ committee finally approved the next C++ standard -- commonly referred to as C++11. This new standard defines a multi-threaded abstract machine, together with a well-defined memory model and library support for interaction between threads. The memory model defines when multiple threads may access the same memory location, and specifies when updates by one thread become visible to other threads. This not only allows development of multi-threaded applications in a platform independent and portable way, but also to formally prove the correctness based on the memory order semantics.

Unfortunately at the beginning of the development of Pheet the C++11 standard was not finalized and no implementation of the proposed memory model was available in gcc. For this reason none of the early data structures were developed using the new memory model, but instead with a few macros that resolve to compiler specific intrinsics for atomic operations like fetch-and-add or compare-and-swap.
My task was to provide an adapted implementation of the CentralKStrategy data structures based on the new memory model instead of the old macros and to provide a reasoning about the correctness of the new implementation based on the memory order semantics.

The following description of the implementation and the code listings are based on revision 573 from \url{https://launchpad.net/pheet}.

\section*{The C++11 memory model\footnote{Since the official C++ standard is not freely available I will instead refer to the ''Working Draft, Standard for Programming Language C++`` from January 2012 \cite{c++11_standard} which contains the C++11 standard plus minor editorial changes.}}\label{C++11 memory model}
The new memory model of C++11 is largely based on the work by Boehm, Alexandrescu et al \cite{Boehm:2008:FCC:1379022.1375591, Alexandrescu:2004}.

One of the most important aspects is the definition of a data race \cite[p. 14]{c++11_standard}:
\begin{quote}
The execution of a program contains a data race if it contains two conflicting actions in different threads, at least one of which is not atomic, and neither happens before the other. Any such data race results in undefined behavior.
\end{quote}
Where conflicting actions are defined as follows \cite[p. 11]{c++11_standard}:
\begin{quote}
Two expression evaluations conflict if one of them modifies a memory location and the other one accesses or modifies the same memory location.
\end{quote}
This definition implies that any program written according to the old standard that uses some other threading libraries and shares any data between those threads exhibits undefined behavior.
The memory operations are ordered by means of the \emph{happens-before} relationship that can be roughly described as follows:
\begin{quote}
Let A and B represent operations performed by a multi-threaded process. If A happens-before B, then the memory effects of A effectively become visible to the thread performing B before B is performed.
\end{quote}
The \emph{happens-before} relation (denote: $\to$) is a strict partial order and as such it is transitive, irreflexive and antisymmetric.
\begin{description}
	\leftskip2em
	\item[transitivity] - $\forall a, b, c, \textrm{if } a \to b \textrm{ and } b \to c, \textrm{ then } a \to c$
    \item[irreflexivity] - $\forall a, a \not\to a$
    \item[antisymmetry] - $\forall a, b, \textrm{if } a \to b\ \textrm{ then } b \not\to a$
\end{description}

The complete formal definition specifically for C++ can be found in \cite[p. 11-13]{c++11_standard}.

A \emph{happens-before} order between two operations from the same thread (program order) is implicitly given by the \emph{sequenced-before} order \cite[p. 10,13]{c++11_standard}. A \emph{happens-before} order between two operations from different threads (in the standard this is referred to as \emph{inter-thread-happens-before}) must be established using atomic operations.

The following memory orders can be specified for each atomic operation (from strong to relaxed):
\begin{itemize}
	\item \texttt{memory\_order\_seq\_cst}
	\item \texttt{memory\_order\_acq\_rel}
	\item \texttt{memory\_order\_release}
	\item \texttt{memory\_order\_acquire}
	\item \texttt{memory\_order\_consume}
	\item \texttt{memory\_order\_relaxed}
\end{itemize}

\texttt{memory\_order\_consume} and \texttt{memory\_order\_acquire} can only be used for operations that perform a \textit{read}, \texttt{memory\_order\_release} can only be used for operations that perform a \textit{write} and \texttt{memory\_order\_acq\_rel} can only be used for operations that perform a \textit{read-modify-write} operation.
Although the language does not enforce these constraints some implementations do check them at runtime\footnote{For example the Microsoft STL implementation verifies these constraints at runtime when the \texttt{DEBUG} macro is defined.}.

A \emph{happens-before} relationship can be established by using the following combinations of memory orders\footnote{\texttt{memory\_order\_acq\_rel} is the combination of \texttt{memory\_order\_release}
and \texttt{memory\_order\_acquire}. So wherever either one is used it is also possible to use \texttt{memory\_order\_acq\_rel}.}:
\begin{itemize}
	\item \texttt{memory\_order\_seq\_cst + memory\_order\_seq\_cst}
	\item \texttt{memory\_order\_acquire + memory\_order\_release}
	\item \texttt{memory\_order\_consume + memory\_order\_release}
\end{itemize}
There is always a \emph{happens-before} order between two \texttt{memory\_order\_seq\_cst} operations \cite[p. 1100]{c++11_standard}.

An atomic operation $A$ that performs a store-release operation on an atomic object $M$ \emph{synchronizes with} an atomic operation $B$ that performs a load-acquire operation on $M$ and takes its value from any side effect in the release sequence (defined below) headed by $A$.
This \emph{synchronize-with} order is compatible with the \emph{inter-thread-happens-before} order.
The \emph{dependency-ordered-before} relation resulting from \texttt{memory\_order\_consume} operations is more complicated, but since it is not used in the ported implementation I will not go into more details (these can be found in \cite[p. 12]{c++11_standard}).
\texttt{memory\_order\_relaxed} can never be used to create a \emph{happens-before} order.

All modifications to a particular atomic object occur in some particular total order, called the \emph{modification order}. If $A$ and $B$ are modifications of an atomic object $M$ and $A$ \emph{happens-before} $B$, then $A$ precedes $B$ in the modification order of $M$. There are separate modification orders for each atomic object and there is no requirement that these can be combined into a single total order for all objects.

A \emph{release sequence} is a subsequence of the modification order of an atomic object. It is headed by a release operation $A$ and followed by an arbitrary number of
\begin{itemize}
	\item atomic operations performed by the same thread that performed $A$ or
	\item atomic read-modify-write operations.
\end{itemize}
For operations performed by the same thread that performed $A$ it is not relevant which memory order is used -- even \texttt{memory\_order\_relaxed}.

The standard describes two different compare-and-swap operations for atomic objects - \texttt{compare\_exchange\_strong} and \texttt{compare\_exchange\_weak}. The difference is that \texttt{compare\_exchange\_weak} is allowed to fail spuriously, that is, act as if \texttt{*obj != *expected} even if they are equal, but can result in better performance on some platforms.


\section*{CentralKStrategy}
CentralKStrategy is a task storage data structure that is internally used by the Pheet scheduler. It creates a global priority ordering between all the tasks available in the system, while allowing each worker thread to miss up to $k$ of the newest tasks, as long as each task is seen by at least one thread. This $k$-relaxation, which is based on the scheme introduced by Afek et al. \cite{Afek:2010:QRC:1940234.1940273}, is used to improve scalability of the data structure because it is shared by all threads.

The implementation is split into two components. The first is a global shared array that is used to share tasks between all threads and to maintain information about which tasks have to be globally visible, to avoid violations of the $k$-relaxation guarantees. The second is a local priority queue for each thread (in Pheet also referred to as ''place``) that is used to maintain the priority ordering of the task visible to each thread.

A detailed description of the CentralKStrategy data structure can be found in \cite{Wimmer:2014:DST:2555243.2555278} and \cite[pp. 81]{Wimmer:2014:diss}.

The CentralKStrategy implementation consists of the following data structures:
\begin{itemize}
\item \texttt{CentralKStrategyTaskStorageItem}
\item \texttt{CentralKStrategyTaskStorageImpl}
\item \texttt{CentralKStrategyTaskStoragePlace}
\item \texttt{CentralKStrategyTaskStorageDataBlock}
\end{itemize}

My first task was to identify all the members of these data structures that have to be defined as atomics in order to avoid any data races. Once these members were changed to atomics the implementation would be correct since all the operations on atomics use \texttt{memory\_order\_seq\_cst}. However, from a performance point of view this is far from optimal, so for every single operation on any atomic member I relaxed the used memory order as far as possible, while ensuring that all the required \emph{happens-before} relations are kept in place.


\subsection*{CentralKStrategyTaskStorageItem}
This data structure represents an item that is referenced from the global array as well as from the thread's local priority queue. Each item represents a generated task to be executed by some worker thread. It has the following members:
\begin{description}
	\leftskip2em
	\item[\texttt{Pheet::Scheduler::BaseStrategy* strategy}] - a pointer to a \textit{scheduling strategy} that is used internally by the scheduler. More details about how the scheduler can be configured with strategies can be found in \cite{Wimmer:2013:WCS:2442516.2442562}.
	\item[\texttt{TT data}] - some implementation specific data.
	\item[\texttt{size\_t position}] - the position of this item in the global array, also used to mark this item as ''taken``.
	\item[\texttt{size\_t orig\_position}] - the position in the global array where this item was originally inserted.
	\item[\texttt{Place* owner}] - pointer to the \texttt{CentralKStrategyTaskStoragePlace} instance that owns this item.
	\item[\texttt{void (Place::*item\_push)(Self* item, size\_t position)}] - pointer to a member function that is used to push the item to a thread's local priority queue.
\end{description}
It does not contain any relevant methods.

The only member that can participate in a data race is \texttt{position}. All other members are set by the owning thread before the item is made globally available (by storing a reference in the global array) and the other threads only read these members. Therefore it is sufficient to make the member \texttt{position} atomic.

\paragraph{position} \hfill \\
\texttt{position} is set to the same value as \texttt{orig\_position} during initialization (together with all the other members) before the item is made globally visible. All other accesses are either loads followed by a \texttt{compare\_exchange} operation to update \texttt{position} and mark the item as ''taken`` (\texttt{position != orig\_position}), or loads followed by a comparison to check whether the item is still active.

In either case a load that ''sees`` an old value is not an issue -- either the following \texttt{compare\_exchange} operation fails or the item is still seen as active which may result in some additional work (e.g. the item is inserted in the thread local heap structure), but does not affect correctness. Since there are no interdependencies with other values there is no need for a \emph{synchronize-with} relation between any loads and stores, so all operations can use \texttt{memory\_order\_relaxed}.


\subsection*{CentralKStrategyTaskStorageImpl}
This data structure represents the global shared array. There exists a single instance that is shared by all places (threads). It has the following members:
\begin{description}
	\leftskip2em
	\item[\texttt{size\_t tail}] - the global tail index in the global array -- \texttt{head} indexes are thread local. It is guaranteed that for every thread all the items between the threads local \texttt{head} and the global \texttt{tail} are filled with items.
	\item[\texttt{DataBlock* start\_block}] - pointer to the first data block of the global array (an instance of \texttt{CentralKStrategyTaskStorageDataBlock}).
	\item[\texttt{size\_t num\_places}] - the total number of places (threads) that are using this data structure (must be known in advance).
\end{description}
It does not contain any relevant methods, most of the methods are implemented in \texttt{CentralKStrategyTaskStoragePlace}. 

\texttt{tail} can be accessed concurrently by several threads and therefore has to be defined as atomic. \texttt{start\_block} and \texttt{num\_places} are only set during initialization and therefore cannot participate in any data race.

\paragraph{tail} \hfill \\
While every thread has its own local \texttt{head} variable (see description of the data structure \texttt{CentralKStrategyTaskStoragePlace}) there is only a single global tail.
It is guaranteed that all the entries between some \texttt{head} and the global \texttt{tail} contain valid pointers to items. Entries are inserted behind \texttt{tail} according to the $k$ relaxation. As soon as all $k$ entries behind \texttt{tail} are used up, \texttt{tail} is updated accordingly.

In order to guarantee a \emph{happens-before} order for load/store accesses to the entries the \texttt{compare\_exchange\_weak} operation on \texttt{tail} in \texttt{push} must use \texttt{memory\_order\_release} and the load operation in \texttt{update\_heap} must use \texttt{memory\_order\_acquire}. Both methods are implemented in \texttt{CentralKStrategyTaskStoragePlace}.

It is sufficient to use \texttt{compare\_exchange\_weak} for the update since it is executed in a loop until either this or some other thread is successful, so a spurious failure does not cause any problem (see Listing \ref{lst:tail_update}).

\begin{lstlisting}[caption={Update of \texttt{tail} in \texttt{CentralKStrategyTaskStoragePlace::push}}, label=lst:tail_update]
size_t nold_tail = task_storage->tail.load(std::memory_order_relaxed);
ptrdiff_t diff = static_cast<ptrdiff_t>(cur_tail) - static_cast<ptrdiff_t>(nold_tail);
while (diff > 0)
{
	if (task_storage->tail.compare_exchange_weak(nold_tail, cur_tail, std::memory_order_release, std::memory_order_relaxed))
		break;
	diff = static_cast<ptrdiff_t>(cur_tail) - static_cast<ptrdiff_t>(nold_tail);
}\end{lstlisting}


\subsection*{CentralKStrategyTaskStoragePlace}
This data structure contains all the thread local data -- including the thread local priority queue. It has the following members:
\begin{description}
	\leftskip2em
	\item[\texttt{PerformanceCounters pc}] - a class containing some data structure specific performance counters.
	\item[\texttt{TaskStorage* task\_storage}] - a pointer to the \texttt{CentralKStrategyTaskStorageImpl} singelton instance that is shared by all threads.
	\item[\texttt{StrategyRetriever sr}] - the strategy retriever that is used in the heap (local priority queue).
	\item[\texttt{StrategyHeap heap}] - the thread's local priority queue.
	\item[\texttt{DataBlock* tail\_block}] - pointer to the first data block in the linked list (that builds the global shared array) that this thread knows of.
	\item[\texttt{DataBlock* head\_block}] - pointer to the last data block in the linked list (that builds the global shared array) that this thread knows of.
	\item[\texttt{size\_t head}] - the thread's local head index in the global shared array.
	\item[\texttt{ItemMemoryManager items}] - memory manager for the items created by this thread.
	\item[\texttt{DataBlockMemoryManager data\_blocks}] - memory manager for the data blocks created by this thread.
\end{description}
Since every thread has its own \texttt{CentralKStrategyTaskStoragePlace} instance and does not touch instances from other threads the members of this data structure cannot participate in any data race and are not required to be atomic.


\subsection*{CentralKStrategyTaskStorageDataBlock}
The global array is actually implemented as a linked list of arrays. This data structure represents one such part of the global array. It has the following members:
\begin{description}
	\leftskip2em
	\item[\texttt{Item* data[BlockSize]}] - an array of \texttt{CentralKStrategyTaskStorageItem} pointers. BlockSize is a template parameter and can therefore be tuned.
	\item[\texttt{size\_t offset}] - the index offset for this block (the first element in this block has index \texttt{offset} in the global array).
	\item[\texttt{DataBlock* next}] - pointer to the next data block in the linked list that builds the whole global array.
	\item[\texttt{size\_t active\_threads}] - the number of threads that have not yet discarded this block.
	\item[\texttt{bool active}] - a flag that signals whether this block is currently in use or not.
\end{description}
The only member that cannot participate in a data race is \texttt{offset} since it is only set during the initialization of a data block. All other members can be accessed concurrently and therefore have to be atomics.

\begin{lstlisting}[caption={C++11 code of \texttt{CentralKStrategyTaskStorageDataBlock::deregister}}, label=lst:new_deregister]
void deregister()
{
	size_t old = active_threads.fetch_sub(1, std::memory_order_relaxed);
	if (old == 1)
	{
		// cleanup data items
		// ...
		next.store(nullptr, std::memory_order_relaxed);
		active.store(false, std::memory_order_release);
	}
}
\end{lstlisting}

A block can be reused as soon as all tasks (items) from it have been executed and the local \texttt{head} indexes of all threads point to positions in successor blocks. A kind of reference counting system is used to track whether a block can be reused or not. During initialization of the block \texttt{active} is set to \texttt{true} and \texttt{active\_threads} is set to the number of total threads. When a thread increases its local \texttt{head} index to point to a new block it calls \texttt{deregister} on the previous block. \texttt{deregister} decreases \texttt{active\_threads} to signal that this thread will no longer reference this block (see Listing \ref{lst:new_deregister}). If \texttt{active\_threads} drops to zero all resources allocated by this block are released and \texttt{active} is set to \texttt{false} to signal that this block can now be reused.

\paragraph{active} \hfill \\
\texttt{active} is used to mark whether a block is currently in use (i.e. can be accessed by several threads concurrently) or not. It is accessed in the following methods:
\begin{description}
	\leftskip2em
	\item[\texttt{is\_reusable}] - checks if \texttt{active} is false (using \texttt{memory\_order\_relaxed})
	\item[\texttt{add\_block}] - reads \texttt{active} (using \texttt{memory\_order\_acquire}) and sets \texttt{active} to true (using \texttt{memory\_order\_relaxed})
	\item[\texttt{deregister}] - sets \texttt{active} to false (using \texttt{memory\_order\_release})
\end{description}
\texttt{is\_reusable} (see Listing \ref{lst:is_reusable}) and \texttt{add\_block} are always executed by the thread that owns the data block. \texttt{deregister} can be executed by several threads concurrently, however only the last thread that decrements \texttt{active\_threads} will clean up the data block and eventually set \texttt{active} to false.

\begin{lstlisting}[caption={\texttt{CentralKStrategyTaskStorageDataBlock::is\_reusable}}, label=lst:is_reusable]
bool is_reusable() const
{
	return !active.load(std::memory_order_relaxed);
}
\end{lstlisting}

We need to ensure a \emph{happens-before} order between resetting the items in \texttt{deregister} and subsequent accesses to them when the block is getting reused after \texttt{add\_block} added it to the linked list. The load operation in \texttt{is\_reusable} uses \texttt{memory\_order\_relaxed} to avoid potentially expensive acquire operations in cases where \texttt{active} is still \texttt{true} and the block cannot be reused yet. Instead \texttt{add\_block} starts with a load operation using \texttt{memory\_order\_acquire} (see Listing \ref{lst:add_block}). The load operation is solely required to establish the \emph{synchronize-with} relation with the store in \texttt{deregister}. A \emph{synchronize-with} relation is only established when the acquire load takes the value that has been written by the release store. In \texttt{add\_block} the value returned by the load can safely be ignored since it is guaranteed to be \texttt{false} (the value written in \texttt{deregister}) because \texttt{add\_block} for this block is only called when \texttt{is\_reusable} returned true -- which (as can be seen in Listing \ref{lst:is_reusable}) is only the case when \texttt{active} is \texttt{false}.

\begin{lstlisting}[caption={\texttt{CentralKStrategyTaskStorageDataBlock::add\_block}}, label=lst:add_block]
void add_block(Self* block, size_t num_places)
{
	// establish synchronize-with relationship with
	// active.store from deregister()
	block->active.load(std::memory_order_acquire);
	block->active_threads.store(num_places, std::memory_order_relaxed);
	block->active.store(true, std::memory_order_relaxed);

	Self* pred = this;
	block->offset = pred->offset + BlockSize;
	auto nextBlock = pred->next.load(std::memory_order_relaxed);
	while (nextBlock == nullptr)
	{
		if (pred->next.compare_exchange_weak(nextBlock, block,
				std::memory_order_release,
				std::memory_order_relaxed))
			return;
	}
	// we failed to add the block, but some other thread must have succeeded
	// -> make our block reusable again
	block->active.store(false, std::memory_order_relaxed);
}
\end{lstlisting}

It would also be possible to create the \emph{synchronize-with} relation between \texttt{deregister} and \texttt{is\_reusable} by using an explicit \texttt{atomic\_thread\_fence} with \texttt{memory\_order\_acquire} (see Listing \ref{lst:is_reusable_fence}).

This way the \emph{synchronize-with} relation with the release store in \texttt{deregister} would only be established when \texttt{active} is actually false. Therefore \texttt{deregister} \emph{happens-before} \texttt{is\_reusable} and \texttt{is\_reusable} \emph{happens-before} \texttt{add\_block}.  From the transitivity of the \emph{happens-before} order follows that \texttt{deregister} \emph{happens-before} \texttt{add\_block}. This way the additional load operations could be omitted. Unfortunately I came up with this idea only a while after I finished the implementation.

\begin{lstlisting}[caption={Alternative implementation of \texttt{is\_reusable} using an explicit \texttt{atomic\_thread\_fence}}, label=lst:is_reusable_fence]
bool is_reusable() const
{
	bool result = !active.load(std::memory_order_relaxed);
	if (result)
		std::atomic_thread_fence(std::memory_order_acquire);
	return result;
}
\end{lstlisting}

\paragraph{active\_threads} \hfill \\
\texttt{active\_threads} is initialized with the number of total threads in \texttt{add\_block} and updated using \texttt{fetch\_sub} in \texttt{deregister}. Since this member is only used to track whether there are still any threads that may read from this block, but it is not used to order other accesses, \texttt{memory\_order\_relaxed} can be used for both operations.

\paragraph{next} \hfill \\
The shared global array is implemented as a single-linked list of data blocks that are linked via the \texttt{next} pointer. The update of the \texttt{next} pointer via \texttt{compare\_exchange\_weak} in \texttt{add\_block} must use \texttt{memory\_order\_release} and the according loads in the methods \texttt{push} and \texttt{deregister\_old\_blocks} of \texttt{CentralKStrategyTaskStoragePlace} must use \texttt{memory\_order\_acquire} to establish a \emph{synchronize-with} relation.

The transitivity of \emph{happens-before} guarantees that resetting the values in \texttt{deregister} \emph{happens-before} any load of an entry from a reused data block in \texttt{push} -– established via the \emph{synchronize-with} relation between the load/store pairs of \texttt{active} and \texttt{next}.

It suffices to use \texttt{compare\_exchange\_weak} for the update since it is executed in a loop until either this thread or some other thread is successful, so a spurious failure does not cause any problem (see Listing \ref{lst:add_block}).

\paragraph{data} \hfill \\
For the \texttt{compare\_exchange\_strong} operation on \texttt{data} in the \texttt{put} method it is required to use \texttt{memory\_order\_release} to ensure a \emph{happens-before} relation between initialization and subsequent loads of the members of \texttt{CentralKStrategyTaskStorageItem}.

In this case it is necessary to use \texttt{compare\_exchange\_strong} to avoid spurious failures because the thread tries to update each entry of a block with $k$ entries only once (see Listing \ref{lst:DataBlock_put}). If this attempt fails for all $k$ entries then it is assumed that all entries are already in use and \texttt{tail} is updated accordingly. However, in the face of spurious failures it may happen that the assumption that all entries are already set is no longer correct. This would break the guarantee that all entries between \texttt{tail} and some \texttt{head} contain valid references.

\begin{lstlisting}[caption={\texttt{CentralKStrategyTaskStorageDataBlock::put}}, label=lst:DataBlock_put]
bool put(size_t& cur_tail, Item* item)
{
	const size_t k = item->strategy->get_k();
	size_t array_offset = cur_tail - offset;
	while (array_offset < BlockSize)
	{
		const size_t cur_k = std::min(k, BlockSize - array_offset - 1);
		const size_t to_add = Pheet::template rand_int<size_t>(cur_k);
		const size_t i_limit = to_add + std::min(Tests, cur_k + 1);
		for (size_t i = to_add; i != i_limit; ++i)
		{
			const size_t wrapped_i = i % (cur_k + 1);
			auto& elem = data[array_offset + wrapped_i];
			if (elem.load(std::memory_order_relaxed) == nullptr)
			{
				auto position = cur_tail + wrapped_i;
				item->orig_position = position;
				item->position.store(position, std::memory_order_relaxed);
				Item* null = nullptr;
				if (elem.compare_exchange_strong(null, item,
						std::memory_order_release,
						std::memory_order_relaxed))
					return true;
			}
		}
		cur_tail += cur_k + 1;
		array_offset = cur_tail - offset;
	}
	return false;
}
\end{lstlisting}

\begin{landscape}

\begin{figure}[h]
	\centering
	\includegraphics[width=1.4\textwidth]{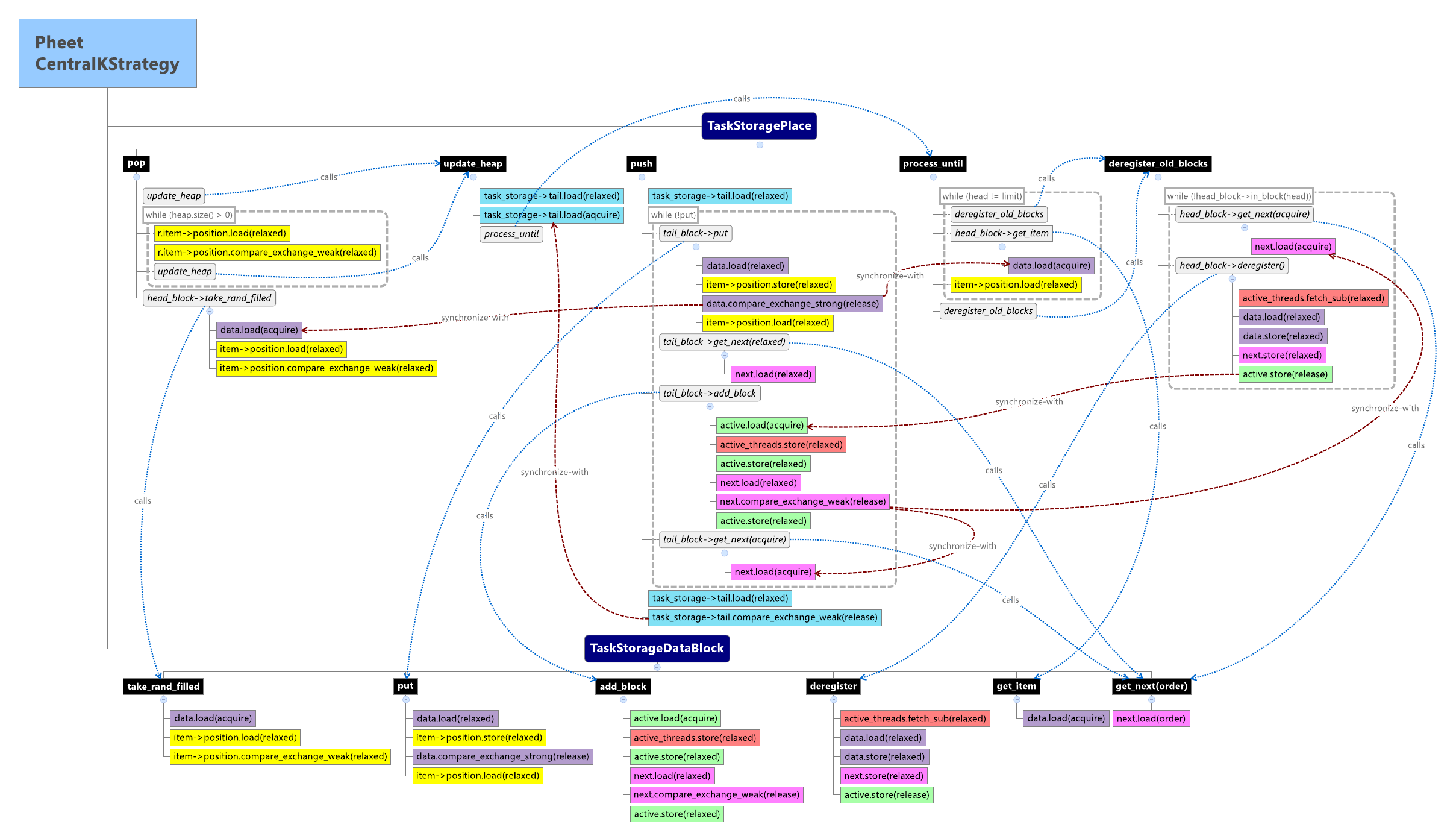}
	\caption{}
	\label{fig:Diagram}
\end{figure}	
\end{landscape}

\hfill \\
The diagram in Figure \ref{fig:Diagram} shows the two classes \texttt{CentralKStrategyTaskStoragePlace} and \texttt{CentralKStrategyTaskStorageDataBlock} together with all their methods that access any atomic variables and calls to other relevant methods. To providea a more complete picture of what is happening the methods of \texttt{CentralKStrategyTaskStorageDataBlock} are ''inlined´´. The different variables are highlighted in different colors so that all the places where they are used can be easily spotted. It also shows the calls between the different methods as well as the established \emph{synchronize-with} relations.


\newgeometry{left=1cm,bottom=1cm,top=1cm,right=1cm}
\restoregeometry


\section*{Benchmarks}
A set of benchmarks have been run to compare the performance of the original implementation against the new C++11 based implementation.
The benchmarks are GP (graph bipartitioning) and SSSP (single source shortest path) from the benchmark suite of the Pheet framework. More information on these and other benchmarks from the Pheet benchmark suite can be found in \cite{Wimmer:2014:diss}.

They were compiled with gcc 4.9.1 and icc 14.0.1 (Intel C++ Compiler) and run on two machines:
\begin{description}
  \item[mars] \hfill \\
		8x Intel\textregistered\ Xeon\textregistered\ E7- 8850 @ 2.00GHz \\
		1TB memory \\
		Linux 3.14-1-amd64 \#1 SMP Debian 3.14.4-1 x86\_64 GNU/Linux
	
  \item[saturn] \hfill \\
		4x AMD Opteron\texttrademark\ 6168 @ 800Mhz \\
		128GB memory \\
		Linux 3.2.0-4-amd64 \#1 SMP Debian 3.2.46-1 x86\_64 GNU/Linux
\end{description}

The benchmarks were run several times with different seeds. The result graphs show the average total runtime.


\subsection*{Graph Bipartitioning}
Graph bipartitioning is a well-known, NP-hard problem \cite{Garey:1990:CIG:574848}. Let $G = (V, E)$ be an undirected, weighted graph, $m = |E|, n = |V|$.
The graph bipartitioning problem is that of partitioning the graph $G$ into two sets with given sizes while minimizing the sum of the weights of the edges that have to be cut.

The Pheet benchmark implements a branch-and-bound approach which is generally well suited to parallelization. It fixes a single node at each step by assigning it to one of the sets for each branch. For bounding (elimination) of sub-problems a simple, easily computable lower bound is used. A more detailed description of this benchmark and its implementation can be found in \cite[pp. 146]{Wimmer:2014:diss}.

The benchmark uses the following parameters:
\begin{description}
	\leftskip3em
	\item[size] the number of nodes in the graph.
	\item[p] controls the density of the graph. The graph's edges are generated randomly using a symmetric adjacency matrix. For each value of the (half) adjacency matrix an edge is added with probability $p$.
	\item[max\_w] the maximum weight of an edge. A uniform random distribution between 1 and $max\_w$ is used to randomly define the weight of an edge.
\end{description}

\begin{samepage}
The test instance used the following parameter values:
\begin{description}
	\leftskip3em
	\item[size] $35$
	\item[p] $0.9$
	\item[max\_w] $1000$
\end{description}
\end{samepage}
It was run 100 times with different seeds.

\begin{figure}[h]
	\centering
	\begin{subfigure}[b]{.49\linewidth}
		\includegraphics{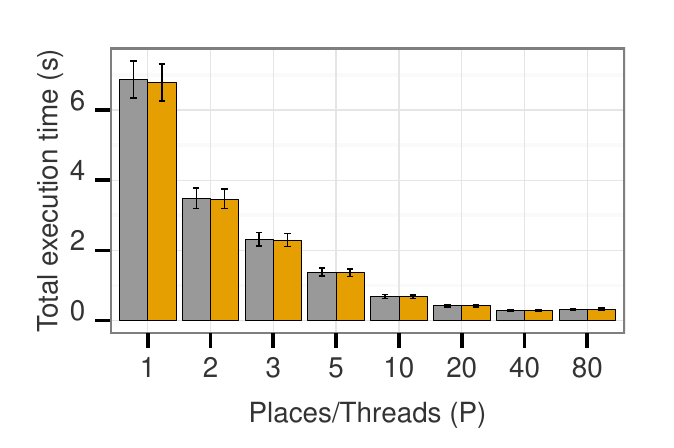}
		\caption{gcc on mars}
	\end{subfigure}
	\begin{subfigure}[b]{.49\linewidth}
		\includegraphics{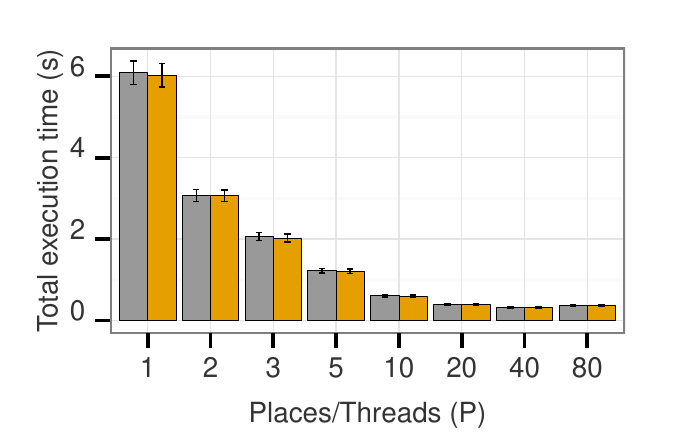}
		\caption{icc on mars}
	\end{subfigure}
	
	\begin{subfigure}[b]{0.49\linewidth}
		\includegraphics{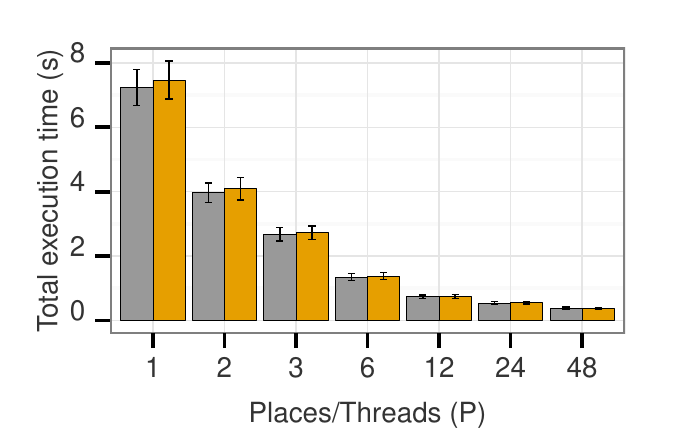}
		\caption{gcc on saturn}
	\end{subfigure}
	\begin{subfigure}[b]{0.49\linewidth}
		\includegraphics{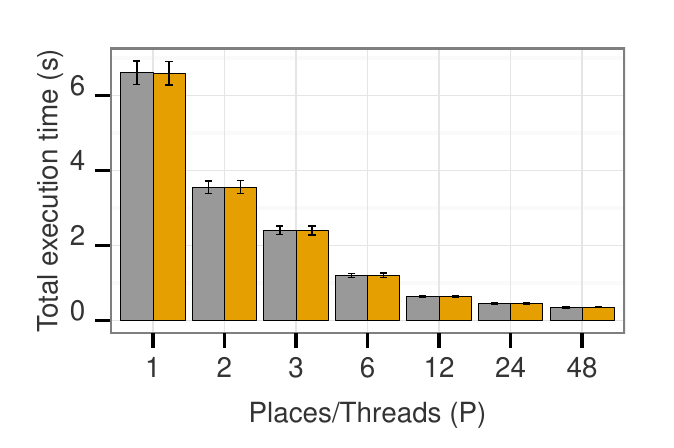}
		\caption{icc on saturn}
	\end{subfigure}
	\begin{subfigure}[b]{0.2\linewidth}
			\includegraphics{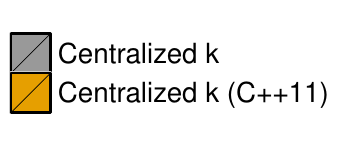}
	\end{subfigure}
	\caption{Graph Bipartitioning benchmark results}
	\label{fig:GP}
\end{figure}

As can be seen from the results in Figure \ref{fig:GP} there is no significant difference in the performance of the two implementations. This is the expected result.
The runtime of the individual tasks is relatively high compared to the runtime overhead of the scheduler. Since I only changed some implementation details of the scheduler, but not the scheduling itself these changes can only impact a small amount of the benchmark's total runtime.


\subsection*{Single Source Shortest Path}
The shortest path problem is a fundamental and well-studied combinatorial optimization problem with many practical and theoretical applications \cite{Tarjan83, Ahuja:1993:NFT:137406}.

Let $G = (V, E)$ be an undirected, weighted graph, $m = |E|, n = |V|$, let $s$ be a distinguished vertex of the graph.

The single source shortest path problem (SSSP) is that of computing, for each vertex $v$ reachable from $s$, a minimum-weight path from $s$ to $v$, where the weight of a path is the sum of the weights of its edges. The Pheet benchmark implementation is based on a simple parallelization of Dijkstra's algorithm. A more detailed description of this benchmark and its implementation can be found in \cite[pp. 164]{Wimmer:2014:diss}.

\begin{figure}[h]
	\centering
	\begin{subfigure}[h]{.49\linewidth}
		\includegraphics{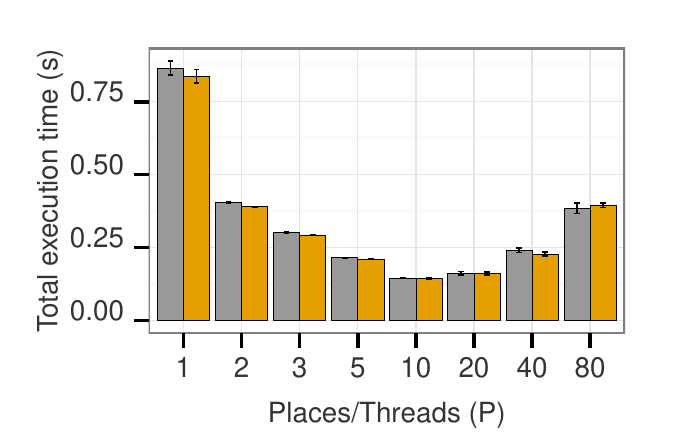}
		\caption{gcc on mars}
	\end{subfigure}
	\begin{subfigure}[h]{.49\linewidth}
		\includegraphics{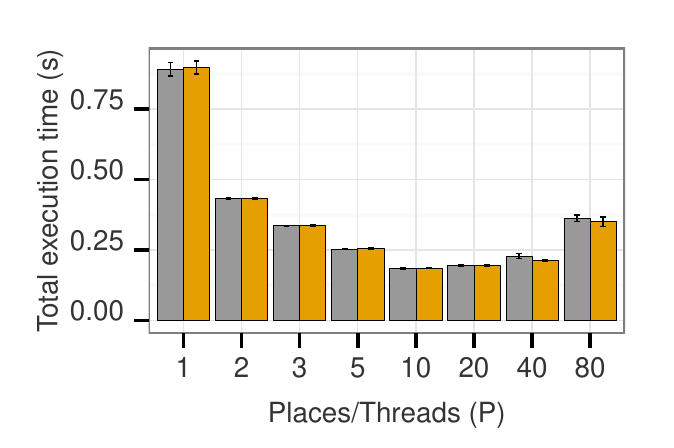}
		\caption{icc on mars}
	\end{subfigure}
	
	\begin{subfigure}[h]{0.49\linewidth}
		\includegraphics{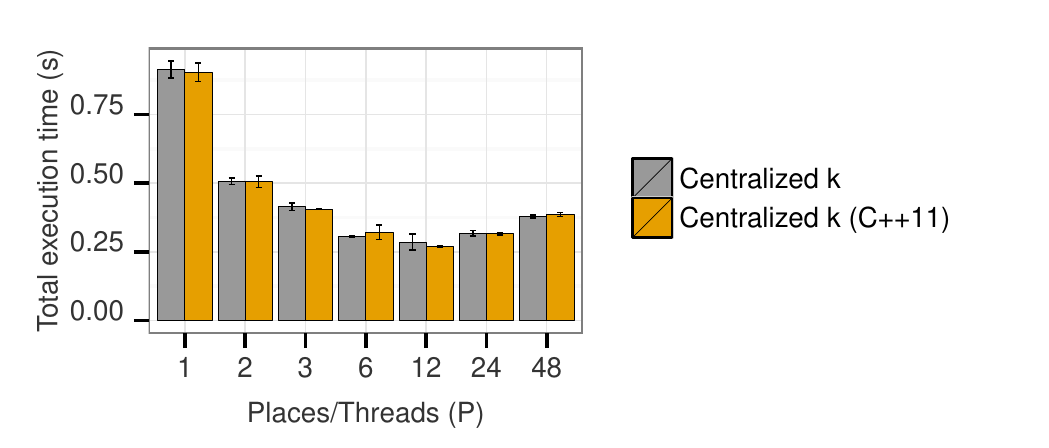}
		\caption{icc on saturn}
	\end{subfigure}
	\caption{Single Source Shortest Path benchmark results for dense graph}
	\label{fig:SSSP_dense}
\end{figure}

The parameters for this benchmark are:
\begin{description}
	\leftskip3em
	\item[size] the number of nodes in the graph.
	\item[p] controls the density of the graph. The graph's edges are generated randomly using a symmetric adjacency matrix. For each value of the (half) adjacency matrix an edge is added with probability $p$.
	\item[max\_w] the maximum weight of an edge. A uniform random distribution between 1 and $max\_w$ is used to randomly define the weight of an edge.
	\item[k] the ''relaxation factor``. This is the number of most recent tasks each thread is allowed to miss.
\end{description}

\pagebreak

\begin{samepage}
The first instance was run 20 times with different seeds using the following parameters (resulting in a dense graph):
\begin{description}
	\leftskip3em
	\item[p] $0.5$
	\item[k] $512$
	\item[max\_w] $100000000$
	\item[size] $10000$
\end{description}
\end{samepage}

The results\footnote{There are not results for gcc on saturn because the machine was needed by someone else so the benchmark could not be run.} in Figure \ref{fig:SSSP_dense} show no significant difference in performance -- just as in the graph bipartitioning benchmark.

\begin{figure}[h]
	\centering
	\begin{subfigure}[b]{.49\linewidth}
		\includegraphics{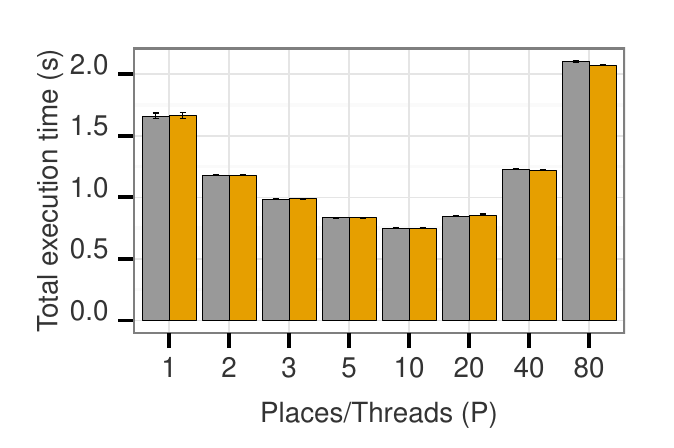}
		\caption{gcc on mars}
	\end{subfigure}	
	\begin{subfigure}[b]{0.49\linewidth}
		\includegraphics{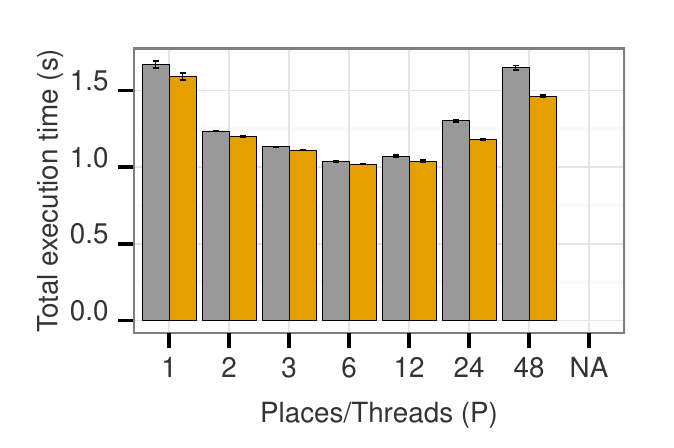}
		\caption{gcc on saturn}
	\end{subfigure}
	\begin{subfigure}[b]{0.2\linewidth}
			\includegraphics{benchmarks/legend.pdf}
	\end{subfigure}
	\caption{Single Source Shortest Path benchmark results for sparse graph}
	\label{fig:SSSP_sparse}
\end{figure}

Because this is a dense graph the generated tasks have high runtime, and therefore again dominate the total runtime. In order to reduce the runtime of the tasks another instance with a sparse graph has been run using the following parameters:
\begin{description}
	\leftskip3em
	\item[p] 0.001
	\item[k] 1024
	\item[max\_w] 100000000
	\item[size] 100000
\end{description}
This instance was run 30 times with different seeds. The results are shown in Figure \ref{fig:SSSP_sparse}.

On Saturn the new C++11 based implementation is significant faster than original implementation, with a $p$-value of \num{2.2e-16} according to a paired wilcoxon signed rank test based on the 210 runs (30 seeds $\times$ 7 thread configurations) per implementation. The values for mean runtime and standard deviation for the Saturn benchmark can be found in Table \ref{tab:SSSP_sparse}.

\begin{table}
\centering
\caption{Mean runtime (s) and standard deviation for sparse graph on saturn.}
\begin{tabular}{r|R{2cm}R{2cm}|R{2cm}R{2cm}}
\toprule
& \multicolumn{2}{c}{\small Centralized k} & \multicolumn{2}{c}{\small Centralized k (C++11)} \\
\cmidrule(lr){2-3} \cmidrule(l){4-5}
Threads & mean & sd & mean & sd \\
\midrule
1 & 1.6713 & 0.0601 & 1.5920 & 0.0597 \\
2 & 1.2357 & 0.0055 & 1.2013 & 0.0069 \\
3 & 1.1343 & 0.0069 & 1.1124 & 0.0048 \\
6 & 1.0373 & 0.0060 & 1.0199 & 0.0065 \\
12 & 1.0714 & 0.0189 & 1.0403 & 0.0185 \\
24 & 1.3032 & 0.0182 & 1.1822 & 0.0125 \\
48 & 1.6493 & 0.0358 & 1.4647 & 0.0180 \\
\bottomrule
\end{tabular}
\label{tab:SSSP_sparse}
\end{table}

The reason for this performance difference most likely is that in the C++11 implementation it is possible to omit an explicit sequential consistent memory fence in \texttt{CentralKStrategyTaskStorageDataBlock::deregister}.

\begin{lstlisting}[caption={Original code of \texttt{CentralKStrategyTaskStorageDataBlock::deregister}}, label=lst:orig_deregister]
void deregister()
{
	size_t old = SIZET_FETCH_AND_SUB(&active_threads, 1);
	if(old == 1)
	{
		// cleanup data items
		// ...

		next = nullptr;
		MEMORY_FENCE(); // <--
		active = false;
	}
}
\end{lstlisting}

In the original code (shown in Listing \ref{lst:orig_deregister}) this explicit memory fence in line 8 was required to ensure that neither the compiler nor the CPU can reorder the two stores. For the new implementation (shown in Listing \ref{lst:new_deregister}) these stores have been replaced by two atomic stores -- the second one using \texttt{memory\_order\_release}. According to the C++11 standard \cite[p. 14]{c++11_standard} the compiler is not allowed to reorder atomic instructions. It is therefore safe to omit the explicit memory fence.

Looking at the results in Figure \ref{fig:SSSP_sparse} it is striking to note that this difference in performance occurs only on Saturn but not on Mars. In the last few years Intel has invested a lot of effort to improve the performance of sequentially consistent operations. The results of the benchmark show that these efforts have really paid off.

Even though it is most likely that the omitting of this memory fence is the cause of the improved scalability I do not have any profiling results that actually prove this assumption.

\section*{Conclusion}
The Pheet CentralK task storage is a good example to show how different concepts of the new C++11 memory model work and why it is great to have a memory model with clearly defined semantics that allows relaxation for every single operation. The whole data structure is implemented using only acquire/release and relaxed semantics, there is no need for any sequential consistent operations. On some architectures the new implementation even performs better than the original.
Making variables atomic and using explicit load/store calls with specific memory order usually produces the same assembler code as the non-atomic operations -- at least on x86/x64 architecture. However, it is absolutely necessary to use atomics for variables that are accessed concurrently by multiple threads to avoid data-races. The point of using the new C++11 memory model is not to improve performance but to make the code both - correct and portable. It is simply the only way to write correct (data-race free) and portable multi-threaded C++ code. The resulting C++ code is not only more expressive (atomic variables and their operations can be easily spotted -- given that the memory model is always defined explicitly), but also allows formal verification of the correctness -- and in some special cases it even results in better performance.

\bibliographystyle{abbrv}
\bibliography{report}

\end{document}